\documentclass[12pt]{article}
\usepackage{amssymb,amsmath}
\usepackage{epsfig}
\usepackage{epstopdf}
\usepackage{latexsym}

\renewcommand{\theequation}{\arabic{section}.\arabic{equation}}

\def\coeff#1#2{\relax{\textstyle {#1 \over #2}}\displaystyle}

\def\IR{\mathbb{R}}
\def\ZZ{\mathbb{Z}}

\def\cF{{\cal F}}

\def\ds{\displaystyle}

\newcommand{\be}{\begin{equation}}
\newcommand{\ee}{\end{equation}}
\newcommand{\bea}{\begin{eqnarray}}
\newcommand{\eea}{\end{eqnarray}}
\def\ep{\epsilon}

\begin{document}

 \begin{titlepage}

\begin{flushright}
IPhT-T08/033
\end{flushright}

\bigskip
\bigskip
\centerline{\Large \bf Spectral Flow, and the Spectrum of
Multi-Center Solutions}
\bigskip
\bigskip
\centerline{{\bf Iosif Bena$^1$, Nikolay Bobev$^2$ and Nicholas P.
Warner$^2$}}
\bigskip
\centerline{$^1$ Institut (Service) de Physique Th\'eorique, }
\centerline{CEA Saclay, 91191 Gif sur Yvette, France}
\bigskip
\centerline{$^2$ Department of Physics and Astronomy}
\centerline{University of Southern California} \centerline{Los
Angeles, CA 90089, USA}
\bigskip
\centerline{{\rm iosif.bena@cea.fr, bobev@usc.edu,
warner@usc.edu} }
\bigskip \bigskip

\begin{abstract}

We discuss ``spectral flow'' coordinate transformations that take
asymptotically four-dimensional solutions into other
asymptotically four-dimensional solutions. We find that spectral
flow can relate smooth three-charge solutions with a multi-center
Taub-NUT base to solutions where one or several Taub-NUT centers
are replaced by two-charge supertubes, and vice versa. We further show that
multi-parameter spectral flows can map such Taub-NUT centers to more
singular centers that are either D2-D0 or pure D0-brane sources.  Since
supertubes can depend on arbitrary functions, we establish
that the moduli space of smooth horizonless black hole microstate
solutions is classically of infinite dimension. We also use the
physics of supertubes to argue that some multi-center solutions
that appear to be bound states  from a four-dimensional
perspective are in fact not bound states when considered from a
five- or six-dimensional perspective.

\end{abstract}

\end{titlepage}

\section{Introduction}

One of the most important problems in quantum gravity is
understanding the origin of black hole entropy. In string theory
one can match this entropy by studying brane configurations at
small or  vanishing string coupling  \cite{stromvafa}, but in this
limit the black hole simply does not exist.  Until recently, very
little was known about what gives rise to the black-hole  entropy
in the regime of parameters where there is, in fact, a classical
black hole. However, the last few years has seen the development
of a programme that addresses this issue and seeks to describe the
black hole entropy in terms of huge number of ``microstate
geometries.''    Such geometries are defined to be smooth,
horizonless backgrounds that have the same charges and asymptotics
as the original black hole. (See \cite{review, Bena:2007kg} for a
review of this proposal.)  Much progress has been made in
constructing such microstate geometries for three-charge black
holes in five dimensions, four-charge black holes in four
dimensions and even for  non-supersymmetric black holes. By now,
it is clear that there exists a huge number of such horizonless
geometries that have the same charges as BPS black holes and black
rings with macroscopically large entropy, and that appear to be
dual to CFT states belonging to the same sector as the CFT states
that give the black hole entropy.

In the quest to understand  black hole entropy in terms of
microstate geometries, two problems appear to be most difficult to
overcome. The first is to determine which of the microstate
solutions are more ``typical'' than others. The second is to
construct very large classes of microstate solutions whose
counting can give the black hole entropy.

Spectral flow has proven to be a  useful tool  in addressing these
kinds of questions.   In the dual conformal field theory the
spectral flow operation is initiated by redefining the $R$-charge
current by mixing it with some other conserved $U(1)$ current.
This then requires a modification of the Hamiltonian in order to
preserve the supersymmetry.  In the bulk gravity theory, the
$U(1)$ $R$-current and the other conserved $U (1)$ current are
dual to isometries of the background and spectral flow can be
achieved simply by a change of coordinates that mixes these two
$U(1)$ directions. One can then add an asymptotically flat region
to this new geometry to obtain a geometry that has different
charges from the original. This is an effective method of obtaining
some five-dimensional three-charge and four-dimensional
four-charge microstate geometries from two-charge geometries
\cite{oleg,gms,BenaKrausKKM,JMaRT}.  In addition, spectral flow
can be used to determine exactly the CFT state dual to the black
hole microstate one constructs, and hence is a useful tool in
determining how typical a certain microstate geometry is.

Despite its usefulness, spectral flow appears to be a rather
cumbersome  operation on asymptotically flat five-dimensional
geometries: One must first strip  the geometry of its
asymptotically-flat region, then perform the spectral flow, and
then add back the aysmptotically-flat geometry. The last step can
be quite non-trivial, especially for geometries that do not have a
large number of isometries (see, for example \cite{Ford:2006yb}).

In this paper we explore a simpler way to use spectral flow to
generate asymptotically four-dimensional geometries starting from
other asmptotically four-dimensional geometries, without stripping
away the asymptotically flat region. This method has two immediate
applications which we believe are quite useful in the programme of
constructing microstates and finding their CFT dual. First, it
allows us to use a known microstate solution to generate a huge
number of other smooth microstate solutions. Secondly, it gives us new insights
into which microstate geometries represent bound states in the CFT.
Since a configuration that consists purely of concentric (two-charge) supertubes
is unbound, any spectral flow of this will give unbound states.  In particular,
we expect such solutions will not  correspond to CFT states in the sector that is
primarily responsible for the entropy.  We will use this observation to examine
the status of some of the microstate geometries that have been studied in the
past.

The fact that one can relate bubbling solutions with a
Gibbons-Hawking (GH), multi-centered Taub-NUT base to solutions
with a supertube in a bubbling solution also indicates that in the
vicinity of the black hole microstates with a GH base there exists
a very large family of other, less symmetric microstate solutions
with the same macroscopic charges.  Indeed, we know from the
Born-Infeld action that two-charge supertubes can have arbitrary
shapes \cite{Mateos:2001qs}, and that these arbitrary shapes
correspond (upon dualizing to the D1-D5-P duality frame) to smooth
geometries \cite{Lunin:2001jy,Lunin:2002iz}.  Hence, one can use
spectral flow to transform a GH center into a supertube, wiggle
the supertube, and undo the spectral flow, to obtain bubbling
three-charge solutions that depend classically on several
arbitrary continuous functions. Hence the dimension of the moduli
space of smooth black hole microstate solutions is classically
infinite.  If, upon counting these solutions, one finds a
black-hole-like entropy, this will be, in our opinion, compelling
evidence that the microstates of black holes are given by
horizonless configurations.  In a forthcoming paper
\cite{entropycount} we will indeed argue that for the deep, smooth
microstate solutions of \cite{Bena:2006kb,abyss} one can obtain an
entropy with the correct charge dependence using the methods
outlined here.

To clarify the relationship of the solutions discussed here with
some earlier results, we note that it was shown in
\cite{5Dsugra,GutowskiYV,BenaDE} that general BPS configurations
with the same supersymmetries as a black hole or black ring
require that the four-dimensional spatial base of the solution be
hyper-K\"ahler. It should be remembered that in establishing this
result it was assumed that the solution was independent of the
``internal'' directions of the compactification tori.  The
solutions that we discuss here, which come from the spectral flow
of supertubes of arbitrary shape, necessarily depend upon one of
these internal directions. Hence, they are more  general than
those considered in \cite{BenaDE}, corresponding to solutions of
ungauged supergravity in six dimensions \cite{Gutowski:2003rg},
and their base space is not hyper-K\"ahler but almost
hyper-K\"ahler.

In section 2 we discuss general five-dimensional BPS solutions,
their relation to solutions of six-dimensional supergravity, and
the way in which the spectral flow transformation acts on a
five-dimensional solution with a $U(1)$ isometry. In section 3 we
specialize this to a translational $U(1)$ isometry where the
solution has a multi-centered Taub-NUT (Gibbons-Hawking) base. We
show that spectral flow acts by interchanging the harmonic
functions underlying these solutions, while keeping the solutions
smooth. The explicit transformation is given in equations
(\ref{GHspecflow1}) and (\ref{GHspecflow}). We also show that
spectral flow is part of a larger $SL(2,\ZZ)^3$ subgroup of the
four-dimensional $E_{7(7)}$ U-duality group, and this particular
subgroup of $E_{7(7)}$ is distinguished because, for generic
parameters, it generates orbits of  smooth solutions.

In section 4 we show that spectral flow can transform a
configuration containing one or several supertubes in Taub-NUT
into a multi-center bubbling solution; conversely, it can
transform such a solution into a solution where at least one of
the centers is replaced by a two-charge supertube. This
demonstrates that the black hole microstates with a GH base
constructed so far in the literature are part of an
infinite-dimensional moduli space of smooth supersymmetric
solutions.  In section 5 we explore the action of generalized
spectral flow on multi-center D6-D4-D2-D0 configurations and use
the physics of supertubes to argue that some multi-center
configurations that appear bound from a four-dimensional
perspective are in fact not bound when seen as full
ten-dimensional solutions. Section 6 contains conclusions, and
Appendix A contains a more detailed discussion of the  $SL(2,\ZZ)^3$ group
of generalized spectral flow transformations.

\section{Three charge solutions}

\subsection{Five-dimensional BPS solutions}

In the M-theory frame, the background that preserves the same
supersymmetries as a BMPV black hole or a supersymmetric black
ring has a metric of the form \cite{GutowskiYV,BenaDE}:
\begin{eqnarray}
 ds_{11}^2  = ds_5^2 & + &    \left(Z_2 Z_3  Z_1^{-2}  \right)^{1\over 3}
 (dx_5^2+dx_6^2) \nonumber \\
 & + & \left( Z_1 Z_3  Z_2^{-2} \right)^{1\over 3} (dx_7^2+dx_8^2)    +
  \left(Z_1 Z_2  Z_3^{-2} \right)^{1\over 3} (dx_9^2+dx_{10}^2) \,.
\label{elevenmetric}
\end{eqnarray}
The five dimensional metric is
\begin{equation}
ds_5^2 ~\equiv~ - \left( Z_1 Z_2  Z_3 \right)^{-{2\over 3}}
(dt+k)^2 + \left( Z_1 Z_2 Z_3\right)^{1\over 3} \,
h_{\mu\nu}dx^{\mu}dx^{\nu} \,, \label{fivemetric}
\end{equation}
and for simplicity we have assumed that the six-dimensional
internal manifold is $T^6$. (In general, this could be any compact
Calabi-Yau three-fold.)  Supersymmetry requires the metric,
$h_{\mu\nu}$, to be hyper-K\"{a}hler.  The solutions
(\ref{elevenmetric}) can be arranged to be either asymptotically
five-dimensional or four-dimensional. The three-form gauge field
decomposes into three vector potentials, $A^{(I)}$, for the
Maxwell fields in the five-dimensional space-time
\begin{equation}
\mathcal{A} ~=~ A^{(1)} \wedge dx_1 \wedge dx_2  ~ + ~ A^{(2)}
\wedge dx_3 \wedge dx_4 ~ + ~ A^{(3)} \wedge dx_5 \wedge dx_6 \,.
\end{equation}
The complete solution is determined by a system of three ``BPS
equations''
\begin{eqnarray}
 \Theta^{(I)}   &=& \star_{4}\Theta^{(I)} \,,  \nonumber \\
\nabla^{2} Z_{I} &=& \ds\frac{1}{2}~ C_{IJK}
\star_{4}(\Theta^{(J)}\wedge\Theta^{(K)})  \,,   \label{BPSeqns} \\
dk +\star_{4}dk &=& Z_{I}~\Theta^{(I)}  \,,  \nonumber
\end{eqnarray}
where we have used the ``dipole field strengths'' $\Theta^{(I)}$
\begin{equation}
\Theta^{(I)} = dA^{(I)} ~+~ d\left(\ds\frac{dt+k}{Z_I}\right)
\end{equation}
and  $\star_{4}$ is the Hodge dual taken with respect to the
four-dimensional hyper-K\"{a}hler metric $h_{\mu\nu}$. The
constants $C_{IJK}$ are the triple intersection numbers of the
Calabi-Yau three-fold, and for $T^6$ we have simply $C_{IJK} =
|~\epsilon_{IJK}|$.

\subsection{Six-dimensional BPS solutions and dimensional reduction}

By dualizing one can recast the foregoing solution in the IIB
frame in which the three fundamental charges are those of the
D1-D5-P system.   In this form, the D5-brane wraps a four-torus,
$T^4$, while the D1-brane, the remaining spatial part of the
D5-brane and the momentum follow a common $S^1$.  The metric thus
naturally decomposes into a six-dimensional part and the
$T^4$-part:
\begin{equation}
 ds_{10}^2  ~=~  ds_6^2 ~+~   Z_1^{1/2}   Z_2^{-{1/2}} \,  ds_{T^4} \,,
\label{tenmetric}
\end{equation}
with
\begin{equation}
ds^2_6 = -\ds\frac{2}{H} (dv+ \beta)\left(du + k + \frac{1}{2}
F(dv+\beta)\right) + H\, h_{\mu\nu}dx^{\mu}dx^{\nu} \label{6Dmet}
\end{equation}
and
\begin{equation}
H ~=~ \sqrt{Z_1 Z_2}  \,,  \qquad F ~=~  -Z_3\,,  \qquad d\beta =
\Theta^{(3)} \,.
\end{equation}
In this formulation there is obviously no longer a symmetry
between the three fundamental charges and, with the foregoing
choices,  $Z_1$ corresponds to the D1-charge,  $Z_2$ to the
D5-charge and  $Z_3$  to the KK-momentum  charge.

We have cast the six-dimensional metric in the form (\ref{6Dmet})
because it affords the easiest comparison with previous work on
the classification of all supersymmetric solutions of
six-dimensional minimal supergravity, obtained in
\cite{Gutowski:2003rg}.   In the minimal theory, two of the $U(1)$
Maxwell fields, $\Theta^{(I)}$,  are set equal and they appear  in
a three-form field strength:
\begin{equation}
G^{(3)} = d(H^{-1}(dv+\beta)\wedge (du+k)) + (dv+\beta)\wedge
\mathcal{G}^{+} + \star_4 dH \label{6Dform}
\end{equation}
with
\begin{equation}
\mathcal{G}^{+}  ~=~  \Theta^{(1)} ~=~ \Theta^{(2)}  \qquad d\beta
= \Theta^{(3)} \,.
\end{equation}
One also has $Z_1= Z_2$.  We will, however, not make this
restriction here\footnote{This corresponds to solutions of
six-dimensional, ungauged supergravity with one tensor multiplet,
and was studied in \cite{Cariglia:2004kk}.} but this earlier work
is of relevance here because it allowed more general backgrounds
that could depend upon the extra background coordinate, $v$.  The
spectral flow operations that we wish to consider could generate
such $v$-dependent solutions. See, for example,
\cite{Ford:2006yb}.

The important point in going to the six-dimensional metric from
the five-dimensional solution is that one of the $U(1)$ gauge
fields has been converted to a six-dimensional Kaluza-Klein field,
$\beta$.  This then puts it on the same footing as a $U(1)$
isometry on the four-dimensional base.  In particular, one can
then mix these two directions with a coordinate transformation and,
as we will see, this generates a spectral flow transformation.
One should also note that one has the freedom to choose which of
the three $U(1)$ Maxwell fields in five dimensions will become the
six-dimensional Kaluza-Klein field and so there are three
independent ways of generating the spectral flow.  We now discuss
this in detail.

\subsection{Spectral Flow}

From the six-dimensional perspective the operation of ``spectral
flow'' is simply a coordinate change that mixes periodic
coordinates on the base with the extra Kaluza-Klein coordinate,
$v$ (see, for example \cite{specflowrefs}). When the base is
asymptotic to $\IR^4$, the size of the circles that are mixed with
the Kaluza-Klein circle becomes infinite, and the spectral flow
operation changes the asymptotics of the solution.  We will bypass
this problem by focusing on solutions that are asymptotically
$\IR^3 \times S^1$.

If the base metric has an isometry then one can adapt the
coordinate system to that isometry and take the metric to be
invariant under translations of a coordinate, $\tau$. In
particular, the base metric can be written in the form:
\be\label{basemet} ds_4^2~=~h_{\mu \nu}dx^\mu dx^\nu ~=~ V^{-1}(d
\tau + A)^2~+~ V \gamma_{ij} dx^i dx^j\,, \ee
where $i,j = 1,2,3$ and every component of the metric is
independent of $\tau$.  The one-form, $A$, and the three-metric,
$\gamma_{ij}$ are, {\it a priori}, arbitrary\footnote{However, the
condition that the base metric be hyper-K\"ahler means that this
metric can be completely determined by solving the $SU(\infty)$
Toda equation \cite{BoyerMM,DasGegenberg,BakasGR,Bena:2007ju}.
This fact will not be needed here.}.

We will also assume that the complete six-dimensional solution is
invariant under $\tau$-translations and for simplicity we will
also assume that the six-dimensional solution is independent of
$v$ but neither of these assumptions is essential to the spectral
flow transformations. It is convenient to decompose the one-forms,
$k$ and $\beta$, according to:
\be\label{basedecomp}{k ~=~ \mu (d \tau + A) ~+~  \omega\,, \qquad
\beta ~=~ \nu (d \tau + A) ~+~  \sigma \,,} \ee
where $\omega$ and $\sigma$ are one-forms in the three-dimensional
space.

A spectral flow is then generated by the change of coordinate:
\be\label{specflow}{\tau ~\to~ \tau ~+~ \gamma \, v\,,} \ee
for some parameter, $\gamma$.  For this to be a well-defined
coordinate transformation on the two circles, $\gamma$ must be
properly quantized\footnote{For a Gibbons-Hawking base $\tau$ has
a period of $4\pi$ and $v$ has a period of $2\pi$ so $\gamma$ has
to be an even integer.}. More generally we could consider any
global diffeomorphism in the $SL(2,\ZZ)$ that acts on the
two-torus defined by these $U(1)$'s. We will return to this in
section 3. The important point is that because these mappings are
diffeomorphisms, they map regular solutions without closed time-like
curves (CTC's)  onto regular solutions without closed time-like  curves.

Inserting (\ref{specflow}) into (\ref{6Dmet}), one can   collect
terms and restore the entire metric back to its canonical form,
(\ref{6Dmet}).  One finds that this coordinate transformation is
equivalent to:
\be\label{modsixmet} ds_6^2 ~\to~ d\tilde s_6^2~\equiv~ -2
\widetilde H^{-1} \, (dv+\tilde \beta) \,  \big(du + \tilde k +
\coeff{1}{2}\, \widetilde \cF\, (dv+\tilde \beta)\big) ~+~
\widetilde H \, d\tilde s_4  \,, \ee
where
\bea \widetilde V ~=~&  (1+ \gamma \, \nu )\, V \,, \qquad
\widetilde A ~=~  A - \gamma \, \sigma\,, \qquad
 \widetilde H ~=~   (1+ \gamma \, \nu)^{-1} H \,,\nonumber \\
\tilde \beta ~=~& (1+ \gamma \, \nu)^{-1} \beta \,, \qquad
\widetilde F~=~   (1+ \gamma \, \nu)\, F +  2 \gamma \mu +
(1+ \gamma \, \nu)^{-1} V^{-1} \gamma^2 H^2\,, \label{transfobjs} \\
\tilde k   ~=~& \displaystyle  k ~-~ {\gamma  \mu  \over (1+
\gamma \, \nu)} \,  \beta+
 {\gamma^2 H^2 \over V\, (1+ \gamma \, \nu)^{2}  }\,  \beta -
{ \gamma  H^2 \over V\, (1+ \gamma \, \nu)} \, (d \tau + A)  \,.
\nonumber \eea

For a general hyper-K\"ahler metric with a rotational $U(1)$
isometry, two of the three complex structures depend explicitly
upon $\tau$ \cite{BakasGR,Bena:2007ju} and so, after the shift
(\ref{specflow}), these two complex structures depend upon $v$. As
a result, the metric, $d\tilde s_4^2$ is almost-hyper-K\"ahler
\cite{Gutowski:2003rg} but not hyper-K\"ahler.  On the other hand,
if the $U(1)$ isometry is translational then the hyper-K\"ahler
metric may be put into Gibbons-Hawking form \cite{GibbonsZT} and
all three complex structures are independent of $\tau$ and so
$d\tilde s_4^2$ will also be hyper-K\"ahler with a translational
$U(1)$ isometry and hence must have Gibbons-Hawking form
\cite{Gibbons:1987sp}. We now investigate this in more detail.

\section{Solutions with a Gibbons-Hawking base}

\subsection{Review of solutions with a Gibbons-Hawking bases}

The Gibbons-Hawking metrics have the form
\begin{equation}
h_{\mu\nu} ~=~ V^{-1}(d\tau + \vec{A} \cdot d\vec{y})^2 + V
(dy_1^2 + dy_2^2 + dy_3^2) \label{GHmet}
\end{equation}
where we write $\vec{y} = (y_1,y_2,y_3)$ and where
\begin{equation}
\vec{\nabla} \times \vec{A} = \vec{\nabla} V \,.
\end{equation}
This means that $V$ must be harmonic on the $\mathbb{R}^3$ spanned
by $\vec{y}$ and one should recall that to avoid orbifold
singularities at singular points of $V$ the coordinate $\tau$ has
to have the range $0\leq \tau \leq 4\pi$. The solutions of
equations (\ref{BPSeqns}) with a Gibbons-Hawking base were derived
in \cite{5Dsugra,Gauntlett:2004qy,Bena:2005va,Berglund:2005vb}

The metric (\ref{GHmet}) has a natural set of frames:
\begin{equation}
\hat{e}^1 = V^{- \frac{1}{2}} (d\tau + A), \qquad \hat{e}^{a+1} =
V^{1\over2}~dy^a, \,\,\,\, a = 1,2,3\,,
\end{equation}
where $A \equiv \vec{A}\cdot d\vec{y} $. There are also two
natural sets of two-forms:
\begin{equation}
\Omega_{\pm}^{(a)} ~\equiv~ \hat{e}^1 \wedge \hat{e}^{a+1} \pm
\ds\frac{1}{2} ~ \epsilon _{abc}~ \hat{e}^{b+1}\wedge
\hat{e}^{c+1}, \qquad a=1,2,3.
\end{equation}
The $\Omega_{-}^{(a)}$ are anti-self-dual and harmonic, defining
the hyper-K\"{a}hler structure on the base. The forms,
$\Omega_{+}^{(a)}$, are self-dual, and we can take the self-dual
field strengths, $\Theta^{(I)}$, to be proportional to them:
\begin{equation}
\Theta^{(I)} ~=~ - \ds\sum_{a=1}^{3}~
(\partial_{a}(V^{-1}K^{I}))~\Omega_{+}^{(a)} \,. \label{Thetadefn}
\end{equation}
For $\Theta^{(I)}$ to be closed, the functions $K^I$ have to be
harmonic on $\mathbb{R}^3$. One can easily find potentials, $B^I$,
with $\Theta^{(I)} = dB^I$:
\begin{equation}
B^{I} = V^{-1} K^{I} (d\tau + A) + \vec{\xi}^{I} \cdot d\vec{y}
\,,
\end{equation}
where
\begin{equation}
\vec{\nabla} \times \vec{\xi}^{I} = - \vec{\nabla} K^I \,.
\end{equation}
Hence, $\vec{\xi}^{I}$ are vector potentials for magnetic
monopoles located at the poles of $K^{I}$. The three self-dual
Maxwell fields $\Theta^{(I)}$ are thus determined by the three
harmonic functions $K^{I}$. Inserting this result in the right
hand side of (\ref{BPSeqns}) we find:
\begin{equation}
Z_{I} = \ds\frac{1}{2} C_{IJK} V^{-1} K^{J} K^{K} ~+~ L_{I}
\end{equation}
where $L_{I}$ are three more independent harmonic functions.

We now write the one-form, $k$, as:
\begin{equation}
k = \mu(d\tau + A) + \omega
\end{equation}
and then the last equation in (\ref{BPSeqns}) implies:
\bea \mu &~=~& \ds\frac{1}{6} C_{IJK}V^{-2}K^{I}K^{J}K^{K} ~+~
\ds\frac{1}{2} V^{-1}K^{I}L_{I} ~+~ M \,, \\
\vec{\nabla} \times \vec{\omega} &~=~& V\vec{\nabla} M ~-~
M\vec{\nabla} V  + \ds\frac{1}{2} (K^I\vec{\nabla} L_I ~-~
L_I\vec{\nabla} K^I)~, \label{omegaeqn} \eea
where $M$ is another harmonic function.

The solution is therefore characterized by the eight\footnote{For
a general $U(1)^N$ five-dimensional ungauged supergravity this
number is $2N+2$.} harmonic functions $V$, $K^{I}$, $L_{I}$ and
$M$. However the   solution is invariant under the following
shifts
\begin{eqnarray}
 V &\to& V, \qquad K^{I} \to K^{I} + c^{I}V  \,, \nonumber \\
L_{I} &\to& L_{I} - C_{IJK}c^{J}K^{K} - \ds\frac{1}{2} C_{IJK}
c^{J}c^{K}V  \,, \label{gaugetransf}\\
M &\to& M - \ds\frac{1}{2}c^{I}L_{I} + \ds\frac{1}{12}C_{IJK}
(c^{I}c^Jc^K V + 3 c^I c^J K^K) \,, \nonumber
\end{eqnarray}
where $c^I$ are three arbitrary constants. As can be seen from
(\ref{Thetadefn}), these shifts do not modify the metric and field
strengths and so should be viewed as a gauge transformation.  One
would therefore expect that any physical quantity will be
invariant under this transformation.  There is, however, a minor
subtlety: when the topology of the base is $\IR^3 \times S^1$, the
Maxwell fields can have Wilson lines around the $S^1$ and then
(\ref{gaugetransf}) can modify the Wilson lines in non-trivial
ways.

The eight harmonic functions that give the solution may be
identified with the eight fundamental basis elements of the
fifty-six dimensional representation of  $E_{7(7)}$:
\begin{eqnarray}
x_{12} &=& L_1, \qquad x_{34} ~=~ L_2, \qquad x_{56} ~=~ L_3,
\qquad x_{78} ~=~ -V \,,
\nonumber\\
y_{12} &=& K^1, \qquad y_{34} ~=~ K^2, \qquad y_{56} ~=~ K^3,
\qquad y_{78} ~=~ 2M \,. \label{Esevenreln}
\end{eqnarray}
With these identifications, one can identify the right-hand side
of (\ref{omegaeqn}) in terms of the symplectic invariant of the
\textbf{56} of $E_{7(7)}$:
\begin{equation}
\vec{\nabla}\times \vec{\omega} = \ds\frac{1}{4}\ds\sum _{A,B}
(y_{AB}\vec{\nabla}x_{AB} ~-~ x_{AB}\vec{\nabla}y_{AB}) \,.
\label{newomega}
\end{equation}
The quartic invariant of the \textbf{56} of $E_{7(7)}$ is
determined by:
\begin{multline}
J_{4} ~=~ -\ds\frac{1}{4} ( x_{12}y^{12} + x_{34}y^{34}
+x_{56}y^{56} +x_{78}y^{78} )^2 - (x_{12}x_{34}x_{56}x_{78} +
y^{12}y^{34}y^{56}y^{78})\\
+ x_{12}x_{34}y^{12}y^{34} + x_{12}x_{56}y^{12}y^{56} +
x_{12}x_{78}y^{12}y^{78} + x_{34}x_{56}y^{34}y^{56} +
x_{34}x_{78}y^{34}y^{78}+ x_{56}x_{78}y^{56}y^{78}\,,
\end{multline}
which is also gauge invariant.  This quantity plays a major role
in determining the horizon area of four-dimensional black holes,
and in formulating a necessary condition for the absence of closed
time-like curves in a given solution
\cite{Berglund:2005vb,Bena:2005ni}.

Finally, we note that there is a natural $SL(2,\ZZ)^4$ subgroup of
the $E_{7(7)}$ duality group in which each $SL(2,\ZZ)$ subgroup
acts simultaneously on four pairs of the form $(x_{AB}, y_{CD})$.
Details will be given in the Appendix, where we will show that
three of these $SL(2,\ZZ)$'s are generated by generalized spectral
flow transformations and generalized electric-magnetic dualities.

\subsection{Spectral flow in Gibbons-Hawking metrics }

Elevating, or ``oxidizing,'' the five-dimensional solution to six
dimensions puts one of the $K^I$'s on the same footing as the
function $V$.  One should therefore expect that the gauge
invariance (\ref{gaugetransf}) should be paralleled by similar
shifts of $V$ by $K^I$ in the six-dimensional solution.  This is
precisely what spectral flow achieves: It is a completely trivial
coordinate change in six dimensions but from the five-dimensional
perspective it significantly modifies the underlying geometry and
Maxwell fields.

To make this more explicit, it is useful to rewrite the
six-dimensional supergravity solution with a GH base (\ref{6Dmet})
as \cite{Gutowski:2003rg}:
\begin{equation}
ds^2_6 = -\ds\frac{F}{H} \left[ dv+ \beta + \ds\frac{1}{F} (du +
k) \right]^{2} + \ds\frac{1}{H F} (du + k)^2  + H\left[
\ds\frac{1}{V} (d\tau + A)^2 + V (dx^2 + dy^2 + dz^2) \right] \,,
\label{6DGH}
\end{equation}
where one should recall that $H=\sqrt{Z_1Z_2}$ and $F = -Z_3$.  As
before we define:
\begin{equation}
k = \mu (d\tau + A) + \omega\,, \qquad \beta = \nu (d\tau + A) +
\sigma \,.
\end{equation}

Starting from   M-theory on $T^6$, one can  choose to dualize to
six dimensions so that any one of the $K^I$ becomes the
Kaluza-Klein potential,  and if we take this to be $K^3$ then one
has:
\begin{equation}
\nu ~=~ V^{-1}\, K^3 \,, \qquad    \vec \nabla K^3 ~=~ -  \vec
\nabla \times \vec \sigma \,. \label{nusigform}
\end{equation}
The spectral flow transformation (\ref{transfobjs}) then
corresponds to:
\bea \widetilde L_3 &~=~&   L_3 ~-~  2\, \gamma\, M \,, \qquad
\widetilde L_2 ~=~   L_2 \,, \qquad
\widetilde L_1 ~=~   L_1  \,, \qquad \nonumber \\
\widetilde K^1&~=~& K^1 ~-~ \gamma \,L_2  \,, \qquad \widetilde
K^2~=~ K^2 ~-~ \gamma \,L_1  \,, \qquad
\widetilde K^3~=~ K^3  \,, \qquad \label{GHspecflow1} \\
\widetilde V &~=~& V +  \gamma \, K^3 \,,  \qquad \widetilde M ~=~
M \,, \qquad \vec{\tilde \omega} ~=~ \vec \omega  \,.  \nonumber
\eea
We can also consider a more general process in which each of the
$K^I$'s is successively chosen to be the special one, and a
spectral flow, with parameter $\gamma_I$, is made.  The result is:
\bea \widetilde L_I &~=~&   L_I ~-~  2\, \gamma_I\, M \,, \qquad
\widetilde M ~=~ M \,, \qquad \vec{\tilde \omega} ~=~ \vec \omega  \,, \nonumber \\
\widetilde K^I&~=~& K^I ~-~
C^{IJK}\,\gamma_J \,L_K ~+~  C^{IJK}\, \gamma_J\, \gamma_K\, M \,,\label{GHspecflow} \\
\widetilde V &~=~& V +  \gamma_I \, K^I - \coeff{1}{2}\, C^{IJK}
\, \gamma_I \, \gamma_J \, L_K  + \coeff{1}{3}\,  C^{IJK}
\gamma_I\, \gamma_J\, \gamma_K \, M \,,  \nonumber \eea
where $C^{IJK} ~\equiv~ C_{IJK} ~\equiv~ |\epsilon_{IJK}|$.  The
fact that $\vec \omega$ remains unchanged follows from the
invariance of the source term in (\ref{newomega}).  By exchanging
$x_{AB} \leftrightarrow y_{CD}$ in (\ref{Esevenreln}), one can map
this transformation onto the gauge transformations
(\ref{gaugetransf}). Indeed, one such inversion can be achieved by
$\tau \leftrightarrow v$ and the gauge transformations
(\ref{gaugetransf}) can be generated by coordinate changes of the
form $v \to v + c \tau $.

We will refer to the transformations (\ref{GHspecflow}) as
``generalized spectral flow.''  Unlike the transformation
(\ref{GHspecflow1}), which transforms smooth six-dimensional
solutions into other smooth six-dimensional solutions, the
generalized spectral flow may, in some instances, transform a
smooth solution to a duality frame in which it is no longer
smooth.  We will discuss this further in section 5 while  in the
remainder of this section we examine the $SL(2,\ZZ)$ actions in
more detail and explicitly verify how $SL(2,\ZZ)$ transformations
preserve regularity.

\subsection{$SL(2,\mathbb{Z})$ transformations of bubbling solutions}

The spatial part of the metric (\ref{6DGH}) may be thought of as a
$T^2$ fibration over $\IR^3$, where  $\tau$ and $v$ define the
$T^2$ fiber. As we have seen, spectral flows are generated by the
coordinate transformation (\ref{specflow}).  Similarly, it follows
directly from (\ref{6DGH}) and  (\ref{nusigform}) that the gauge
transformations (\ref{gaugetransf}) with $c^1 =c^2 =0, c^3 =c$ can
be obtained from the coordinate transformation:
\begin{equation}
v ~\to~ v~+~ c \, \tau\,. \label{gaugecoord}
\end{equation}

More generally, one can make any  $SL(2,\mathbb{Z})$
transformation in the global diffeomorphisms of the $T^2$ defined
by $(\tau, v)$:
\begin{equation}
\left( \begin{array}{c}
\tilde{\tau}\\
2\tilde{v}
\end{array} \right) = \mathcal{M} \left( \begin{array}{c}
\tau\\
2v
\end{array} \right)  =   \left( \begin{array}{cc}
m & n\\ p & q
\end{array} \right) \left( \begin{array}{c}
\tau\\
2v
\end{array} \right)~,
\label{globdiffeos}
\end{equation}
Here $\mathcal{M} \in SL(2,\mathbb{Z})$ and the factors of 2
insure the correct periodicities for the $\tilde{\tau}$ and
$\tilde{v}$ coordinates. Since it is a diffeomorphism, any such
transformation will take smooth (CTC-free) solutions to smooth
(CTC-free) solutions.

If one uses this transformation in (\ref{6DGH}) one can easily
recast the metric back into the same form:
\begin{equation}
d\tilde{s}^2_6 = -\ds\frac{\widetilde{F}}{\widetilde{H}} \left[
d\tilde{v}+ \tilde{\beta} + \ds\frac{1}{\widetilde{F}} (du +
\tilde{k}) \right]^{2} + \ds\frac{1}{\widetilde{H} \widetilde{F}}
(du + \tilde{k})^2 + \widetilde{H}\left[
\ds\frac{1}{\widetilde{V}} (d\tilde{\tau} + \widetilde{A})^2 +
\widetilde{V} (dx^2 + dy^2 + dz^2) \right] \,,
\end{equation}
where
\begin{equation}
\tilde{k} ~ \equiv~   \tilde{\mu} (d\tilde{\tau} + \widetilde{A})
+ \tilde{\omega} \,, \qquad \tilde{\beta} ~ \equiv~ \tilde{\nu}
(d\tilde{\tau} + \widetilde{A}) + \tilde{\sigma}\,,
\end{equation}
and
\begin{equation}
\begin{array}{l}
\widetilde{V} = (m - 2n \nu) V, \qquad \widetilde{H} =
\ds\frac{H}{m - 2n\nu}, \qquad \widetilde{F} = (m - 2n\nu) F -
4n\mu - 4n^2 \ds\frac{H^2}{(m - 2n
\nu) V} \,, \\\\
\tilde{\nu} = - \ds\frac{\frac{p}{2} - q\nu}{m - 2n\nu},
\qquad\qquad \tilde{\mu} = \ds\frac{1}{m - 2n\nu} \left( \mu + 2n
\ds\frac{H^2}{(m
- 2n \nu)V} \right) \,,  \\\\
\widetilde{A} = m A + 2n \sigma, \qquad\qquad \tilde{\sigma} = q
\sigma + \frac{p}{2} A, \qquad\qquad \tilde{\omega} = \omega.
\end{array}
\end{equation}

The effect of this $SL(2,\ZZ)$ transformation on the functions
determining the underlying five-dimensional solutions is:
\begin{equation}
\begin{array}{l}
\widetilde{V} = (m - 2n \nu) V, \qquad \tilde{\mu} =
\ds\frac{V}{\widetilde{V}}
\left( \mu + 2n \ds\frac{Z_1 Z_2 }{\widetilde{V}} \right) \,, \\\\
\widetilde{Z}_1 = \ds\frac{V}{\widetilde{V}} Z_1, \qquad
\widetilde{Z}_2 = \ds\frac{V}{\widetilde{V}} Z_2, \qquad
\widetilde{Z}_3 = \ds\frac{\widetilde{V}}{V} Z_3 + 4 n \mu + 4n^2
\ds\frac{Z_1 Z_2 }{\widetilde{V}}\,. \label{VZtrf}
\end{array}
\end{equation}
Note that because the functions $Z_I$ are gauge invariant, their
transformations only depend upon the spectral flow parameter,
$\gamma =-2n$.

Upon identifying the harmonic functions $V$, $K^{I}$, $L^{I}$ and
$M$ that give the solution with the eight $E_{7(7)}$ parameters
$x$ and $y$ (\ref{Esevenreln}), the $SL(2,\ZZ)$ transformation
becomes simply
\begin{equation}
\begin{array}{l}
\left( \begin{array}{c}
\tilde{y}_{12}\\
2\tilde{x}_{34}
\end{array} \right) = \mathcal{M} \left( \begin{array}{c}
y_{12}\\
2x_{34}
\end{array} \right), \qquad\qquad \left( \begin{array}{c}
\tilde{y}_{34}\\
2\tilde{x}_{12}
\end{array} \right) = \mathcal{M} \left( \begin{array}{c}
y_{34}\\
2x_{12}
\end{array} \right)\\  \\
\left( \begin{array}{c}
\tilde{x}_{56}\\
2\tilde{y}_{78}
\end{array} \right) = \mathcal{M} \left( \begin{array}{c}
x_{56}\\
2y_{78}
\end{array} \right), \qquad\qquad \left( \begin{array}{c}
\tilde{x}_{78}\\
2\tilde{y}_{56}
\end{array} \right) = \mathcal{M} \left( \begin{array}{c}
x_{78}\\
2y_{56}
\end{array} \right)
\end{array}
\label{sl2zxy}
\end{equation}

From the point of view of the five-dimensional solution, the
transformation (\ref{sl2zxy}) is simply   a subgroup of the
$E_{7(7)}(\ZZ)$ duality group that takes solutions into solutions.
Nevertheless, the important feature of this transformation is that
it takes smooth solutions into smooth solutions. As we will
discuss below, for generic parameters, (\ref{sl2zxy}) transforms
bubbling solutions into bubbling solutions, while for specific
parameters it can transform them into bubbling solutions that
contain one or several two-charge supertubes, with charges
corresponding to $x_{12}$ and $x_{34}$. As we will see, these
solutions are smooth in the six-dimensional duality frame
(\ref{6Dmet}), but not in five-dimensions.

In order to arrive at the foregoing transformation we chose to
dualize using the function $K^3$ to get the six-dimensional
background.   One can obviously use the other two functions, $K^1$
and $K^2$ and obtain two other $SL(2,\ZZ)$ subgroups of
$E_{7(7)}(\ZZ)$.  Indeed, these three $SL(2,\ZZ)$'s commute with
one another and thus form an   $SL(2,\ZZ)^3$ subgroup of
$E_{7(7)}(\ZZ)$. As could be expected this general $SL(2,\ZZ)^3$
transformation leaves the quartic invariant $J_4$ unchanged. We
discuss this further in the Appendix, where we give the explicit
forms of these transformations.

\subsection{Regularity and the Bubble Equations}

Suppose that the harmonic functions take their usual form for an
ambi-polar Gibbons-Hawking base
\begin{eqnarray}
V &=& \varepsilon_0 ~+~ \ds\sum_{j=1}^{N}\ds\frac{q_j}{r_j}~,
\qquad K^{I} ~=~ k_0^I ~+~ \ds\sum_{j=1}^{N}\ds\frac{k_j^I}{r_j}~,
\nonumber\\\\
L^{I} &=& l_0^I ~+~ \ds\sum_{j=1}^{N}\ds\frac{l_j^I}{r_j}~,\qquad
M~=~ m_0 ~+~ \ds\sum_{j=1}^{N}\ds\frac{m_j}{r_j}~,\label{Fnforms}
\end{eqnarray}
where $r_{j} \equiv |\vec{y} - \vec{y}^{(j)}|$ and
$\varepsilon_0$, $q_j$, $k_a^I$, $l_a^I$, $m_a$ ($a=0,1,\ldots,N$)
are, as yet, arbitrary constants. As usual define:
\begin{equation}
q_0 ~\equiv~ \ds\sum_{j=1}^{N}q_j, \qquad \tilde{k}_j^I ~\equiv~
k_j^I - q_0^{-1} q_j \ds\sum_{j=1}^{N}k_j^I, \qquad \Pi_{ij}^{(I)} ~\equiv~
\left( \ds\frac{k_j^I}{q_j} ~-~ \ds\frac{k_i^I}{q_i}  \right)\,.
\label{chgredefs}
\end{equation}
Recall that for the functions $Z_I$ and $\mu$ to be regular as
$r_j \to 0$, one must take:
\begin{equation}
l_j^I ~=~ -\ds\frac{1}{2}\, C_{IJK}\ds\frac{k_j^Jk_j^K}{q_j},
\qquad m_j ~=~ \ds\frac{1}{12}\,
C_{IJK}\ds\frac{k_j^Ik_j^Jk_j^K}{q_j^2}, \qquad j=1,\ldots,N\,.
\label{chgchoice}
\end{equation}
The constant terms, $\varepsilon_0 $, $k^I_0$, $l^I_0$ and $m_0$,
determine the asymptotic behaviour of the solution.    The
original M-theory geometry is generically asymptotic to $\IR^{1,3}
\times S^1 \times (T^2)^3$ and the constant terms determine the
scales of the $S^1$ and $T^2$ factors and fix the $U(1)$ Wilson
lines around the $S^1$ \cite{Bena:2005ni}.  If one tunes the
constants appropriately ({\it e.g.} if one sets $\varepsilon_0 =0
$) then various circles in the five-dimensional or six-dimensional
metrics will decompactify.

To remove closed time-like curves in the neighborhood of
the points where $r_j\to 0$ one must impose that $\mu\to 0$ as
$r_j \to 0$. Explicitly this yields the bubble equations:
\begin{equation}
\ds\frac{1}{6}C_{IJK} \ds\sum_{j=1,\,j\neq i}^{N} \Pi_{ij}^{(I)}
\Pi_{ij}^{(J)} \Pi_{ij}^{(K)} \ds\frac{q_iq_j}{r_{ij}} ~=~ 2
(\varepsilon_0 m_i - m_0 q_i) + \ds\sum_{I=1}^{3} (k_0^I l_i^I -
l_0^Ik_i^I) \label{bubbleeqns}
\end{equation}
for $i =1, \dots, N$, and where $r_{ij} \equiv |\vec y^{(i)} -
\vec y^{(j)}|$. Summing both sides of this equation and using the
skew-symmetry of $ \Pi^{(I)}_{ij}$ leads to:
\be \label{mzerodefn} {m_0  ~=~  q_0^{-1} \, \Big(\varepsilon_0 \,
m_i   - {1 \over 2} \sum_{j=1}^N\, \sum_I  \big( l^I_0\, k_j^I -
k^I_0\, \ell_j^I  \big) \Big)\,,} \ee
where $q_0$ is given by (\ref{chgredefs}).

We expect that both the regularity of the six-dimensional solution
and the bubble equations are preserved under the simple spectral
flow (\ref{GHspecflow1}) and, more generally, under the global
diffeomorphisms (\ref{globdiffeos}) precisely because they are
diffeomorphisms of the torus.    Moreover, these diffeomorphisms
only involve the space-like sections of the metric and hence they
should not introduce new  CTC's.  One can see this explicitly from
(\ref{VZtrf}). Suppose that $n$ is generic so that $\widetilde V$
and $V$ have {\it exactly the same} singular points.  Then $V^{\pm
1} \widetilde V^{\mp 1}$ is regular and so if one starts with
regular $Z_I$ and $\mu$ then one will end up with regular
$\widetilde Z_I$ and $\tilde \mu$.  Moreover, if the bubble
equations are satisfied then $\mu \to 0$ as $r_j \to 0$ and hence
$\tilde \mu \to 0$ as $r_j \to 0$.  Thus the bubble equations are
satisfied in the new solution.

This argument obviously generalizes to any combination of
transformations in  $SL(2,\ZZ)^3$ that do not change the singular
structure of $V$. Therefore such transformations clearly map
smooth bubbling solutions into smooth bubbling solutions and
preserve the bubble equations.

If the spectral flow parameter, $n$ is not generic, then $V$ and
$\widetilde V$ can have different sets of singular points, but the
solution generated by the simple spectral flow will still be
smooth in six dimensions, and its physics is the subject of the
next section. It turns out that this feature does not generalize
to non-generic many-parameter spectral flow transformations
(\ref{GHspecflow}). These flows will take multi-center black hole
solutions into other multi-center solutions, by preserving the
bubble equations and not introducing closed timelike curves.
However, they may transform microstate solutions that are smooth
in supergravity into solutions that do not appear smooth in
supergravity. This will be the subject of section 5.

\section{Two-Charge Supertubes and Spectral Flow}

Perhaps the most physically interesting spectral flow
transformation occurs when $V$ and $\widetilde V$ have different
sets of singular points.  Suppose that we start with a regular,
bubbled solution and that we use the simple spectral flow
(\ref{GHspecflow1}) so that $\widetilde V$ has (at least) one less
singularity than $V$. It follows that $\widetilde Z_1$,
$\widetilde Z_2$ and $\tilde \mu$ now develop singularities, but
these singularities have a very special form. As we will show,
these singularities correspond exactly to having a two-charge
supertube at the location of the old pole (or poles) of $V$. Going
in the opposite direction, one can start from a geometry
containing one or several two-charge supertubes and obtain a
bubbling solution by doing the inverse spectral flow\footnote{This
is
  exactly the way in which the first three-charge microstates were
  obtained by Lunin, \cite{oleg} and independently by Giusto, Mathur,
  and Saxena \cite{gms}.}.

It is well known that two-charge supertubes give smooth
supergravity solutions when in the duality frame in which they
have D1 and D5 charges and KKM dipole charge, both in flat space
\cite{Lunin:2001jy,Lunin:2002iz} and in Taub-NUT
\cite{BenaKrausKKM}. Since the standard regularity conditions only
involve the local geometry around the supertube, one would expect
two-charge  supertubes to be regular in more generic three-charge
backgrounds \cite{inprogress}. Hence, the fact that the spectral
flow transformation (\ref{GHspecflow}) takes smooth solutions into
smooth solutions is not surprising; after all, from a
six-dimensional perspective, the flow (\ref{GHspecflow}) is
nothing but a coordinate transformation.

The effect of the spectral flow transformation may, at first,
appear surprising from the geometric perspective of the
four-dimensional base: GH-based solutions are   bubbling
geometries with fluxes threading topologically non-trivial cycles
while supertubes are thought of as rotating supersymmetric
ensembles of branes that do not involve topology. The spectral
flow maps one picture into the other and, once again, from the
six-dimensional perspective it is easy to see how this happens.
Consider the (spatial) $U(1)$ fiber parametrized by $v$ in
(\ref{6Dmet}) over any disk that spans the closed loop of the
supertube. At the supertube the function  $H$ in (\ref{6Dmet})
becomes singular and pinches-off the $U(1)$ fiber. The result is a
topologically non-trivial $3$-sphere and the three-form,
(\ref{6Dform}), has a non-zero flux through this $3$-cycle. In the
metric with a GH base, this $3$-cycle simply appears as a
non-trivial $U(1)$ fibration (parametrized by $v$) over a non-trivial
$2$-cycle in the base. The spectral flow merely ``undoes'' the
topology in the base at the cost of introducing an apparent
singularity but both perspectives are equivalent, and describe the
same, completely regular, six-dimensional solution.

\subsection{Bubbling Geometries from Supertubes - One Supertube in
  Taub-NUT}

It is useful to begin by illustrating the spectral-flow procedure
on the solution corresponding to one supertube in Taub-NUT
\cite{BenaKrausKKM}.  The smooth six-dimensional solution
describing two-charge supertubes can be written as a solution with
a GH base using the following harmonic functions
\cite{Bena:2005ni}:
\bea \label{harmonicsuper} V&=& \epsilon_0 + {1\over r}~,~~~L_1 =
1+{Q_1 \over 4 |\vec r - \vec R |}~,~~~
L_2=1+{Q_2 \over 4 |\vec r - \vec R  |}~,~~~L_3= 1\,, \\
K_1& =& 0~,~~~K_2 = 0~,~~~ K_3 =  - {q_3\over 2 |\vec r - \vec R
|} ~,~~~ M = {J_T\over 16} \left({1 \over R} - {1 \over |\vec r -
\vec R  |} \right)~. \eea
where $\vec R$ defines the position of a round supertube that is
wrapping the fiber of the Taub-NUT metric. Not all the constant
parts in the harmonic functions are independent. The absence of
closed timelike curves requires that
\be
\label{radiussuper} J_T \left(\epsilon_0 + {1\over R}\right) = 4
q_3 \ee

Moreover, in six dimensions the metric constructed using
(\ref{harmonicsuper}) is smooth (up to harmless $\mathbb{Z}_{q_3}$
orbifold singularities)  if \cite{BenaKrausKKM}:
\be \label{regularity} {q_3 J_T = Q_1 Q_2~.} \ee
This condition comes from the requirement that $\omega$ in
(\ref{omegaeqn}) has no Dirac-Misner strings.

Before performing the spectral flow, we should observe that the
harmonic functions above can be shifted using a subset of the
gauge transformation (\ref{gaugetransf}) that preserves
$K_1=K_2=0$ and that sets the sum of the coefficients of the poles
in $K_3$ to be zero:
\bea \label{harmonicgauge} V&=& \epsilon_0 + {1\over r}~,~~~L_1 =
1+{Q_1 \over 4 |\vec r - \vec R|}~,~~~
L_2=1+{Q_2 \over 4 |\vec r - \vec R|}~,~~~L_3= 1~,~~~ K_1 = 0~,~~~\, \\
K_2 &=& 0~,~~~ K_3 = {q_3 \ep_0 \over 2} +  {q_3  \over 2 r} -
{q_3 \over 2 |\vec r - \vec R|} ~,~~~ M = {J_T\over 16} \left({1
\over R} - {1 \over |\vec r -\vec R|} \right)- {q_3 \over 4}~.
\eea
Under a spectral flow with parameter $\gamma_3$ one obtains a new
solution with the harmonic functions:
\bea \label{harmonicnew} V&=& \epsilon_0 \left( 1+ {\gamma_3 q_3
\over 2}\right) +  {1\over r}\left( 1+ {\gamma_3 q_3 \over
2}\right) ~ - {q_3 \gamma_3\over 2 |\vec r - \vec R|}~,~~~
K_1 = -\gamma_3 - {\gamma_3 Q_2 \over 4 |\vec r - \vec R|}~,~~~\\
K_2& =& -\gamma_3 - {\gamma_3 Q_1 \over 4 |\vec r - \vec R|}~,~~~
K_3 = {q_3 \ep_0 \over 2} +  {q_3  \over 2 r} - {q_3 \over 2 |\vec r - \vec R|}\,,  \\
L_1&=&1+{Q_1 \over 4 |\vec r - \vec R|}~,~~~
L_2=1+{Q_2 \over 4 |\vec r - \vec R|}~,~~~\\
L_3&=&1+ {\gamma_3 q_3 \over 2} - {\gamma_3 J_T\over 8} \left({1
\over R} - {1 \over |\vec r - \vec R|} \right) ~,~~~ M =
{J_T\over 16} \left({1 \over R} - {1 \over |\vec r - \vec R|}
\right)- {q_3 \over 4}~. \eea

It is not hard to check that the harmonic functions above satisfy
the condition (\ref{chgchoice}), and hence they give a smooth
three-charge two-centered bubbling solution.  Moreover, the
equation that gives the radius of the supertube in Taub-NUT
(\ref{radiussuper}) becomes exactly the ``bubble equation''
(\ref{bubbleeqns}) governing the two-center bubbling solution.
Hence, a spectral flow transformation can be used to change a
smooth two-charge supertube in six dimensions into a smooth
three-charge bubbling solution. This solution has the same
singular parts as the four-dimensional microstate solution
obtained in \cite{toronto}, but has different constant parts in
the harmonic functions.

Of course, to obtain asymptotically five-dimensional solutions
from other asymptotically flat solutions using spectral flow is a
little more complicated. These solutions must not have any
constant term in the $K_I$ \cite{Bena:2005va,Berglund:2005vb}.
Nevertheless, the solution before the spectral flow necessarily
has all the $Z_I$ (and hence $L_I$) limiting to constant values.
Hence, a spectral flow will necessarily introduce a constant term
in at least one of the $K_I$. The way this problem is usually
remedied \cite{oleg,gms,BenaKrausKKM,Ford:2006yb} is to strip off
the asymptotically-flat region of the solution to obtain an
asymptotically $AdS_3$ geometry, spectral flow this geometry, and
then add back by hand the asymptotically-flat part of the
solution.

On the other hand, by looking at the solutions that have
four-dimensional asymptotics, there is no need to eliminate the
constant terms in the $K_I$ harmonic function. A spectral flow
will simply match two solutions with different values of the
moduli at infinity.

\subsection{Bubbling Geometries from Supertubes - Many Supertubes in Taub-NUT}

We can generalize the foregoing example by starting with a
solution describing $N$ two-charge supertubes in Taub-NUT. The
solution is specified by eight harmonic functions which have the
form
\begin{eqnarray}
V &=& \epsilon_0 ~+~ \ds\frac{1}{r}~, \qquad K^1 ~=~ K^2 ~=~0~,
\qquad K^{3} ~=~ k_0^3 ~-~ \ds\sum_{i=1}^{N}\ds\frac{q_3^i}{2
r_i}~,
\nonumber\\
L_{1} &=& l_0^1 ~+~ \ds\sum_{i=1}^{N}\ds\frac{Q_1^i}{4 \,
r_i}~,\qquad L_{2} ~=~ l_0^2 ~+~
\ds\sum_{i=1}^{N}\ds\frac{Q_2^i}{4 \, r_i}~,\qquad
L_{3} ~=~ l_0^3\\
M &=& m_0 ~-~ \ds\sum_{i=1}^{N}\ds\frac{J_i}{16\, r_i}~,\nonumber
\end{eqnarray}
where $r_i ~=~ |\vec{r} - \vec{r}_i|$ and $\vec{r}_i$ are the
locations of the supertubes in the base space. We will also define
$R_i \equiv |\vec{r}_i|$.

If we choose all $\vec{r}_i$ to lie on the negative $z$ axis (in
GH coordinates) this will correspond to a configuration of $N$
concentric supertubes of ``radius\footnote{This is the distance
from the Taub-NUT center to the supertube as measured in the
three-dimensional base, and not the physical radius of the
supertube.},'' $R_i$. It is clear that the straightforward
generalization of the analysis in \cite{BenaKrausKKM} will imply
that, in the duality frame where the two charges of the supertubes
are D1 and D5 charges, the type IIB supergravity solution will be
smooth if (\ref{regularity}) is satisfied for each center:
\begin{equation}
Q_1^i Q_2^i ~=~ q_3^i J_i \,. \label{supermanyregular}
\end{equation}
These $N$ conditions guarantee that the full metric is completely
regular (again up to $\mathbb{Z}_{q_3^i}$ orbifold singularities).
The solution should be free of CTC's and imposing this condition
at the locations of the supertubes and at the origin of the
four-dimensional base yields $N+1$ equations: $N$ expressions that
give the radius of each supertube and generalize
(\ref{radiussuper}), as well as a relation that  fixes the
parameter $m_0$:
\begin{equation}
\left( \epsilon_0 ~+~ \ds\frac{1}{R_i} \right) J_i ~=~ 4 \, {l_0^3
\, q_3^i}~, \qquad\qquad m_0 ~=~  {1 \over 16} \,
\ds\sum_{i=1}^{N} \,  \frac{J_i}{R_i}   \,. \label{radm0}
\end{equation}
We can use the gauge freedom (\ref{gaugetransf}) to fix a gauge in
which $\ds\sum_{i=1}^{N+1} q_3^i = 0$:
\begin{eqnarray}
V &\to& V, \qquad K^{1} \to K^{1}~, \qquad K^{2} \to K^{2}~,
\qquad K^{3} \to K^{3} + c \,V \,,\nonumber \\
L_1 &\to& L_1 - c \, K^2 ~=~ L_1~, \qquad L_2 \to L_2 - c \, K^1 ~=~ L_2~,
\qquad L_3 \to L_3~, \label{gaugetransf2}\\
M &\to& M - \ds\frac{c}{2}\, L_3\,,\nonumber
\end{eqnarray}
where
\begin{equation}
c ~=~  \ds\sum_{i=1}^{N} \, {q_3^i\over 2} \,.
\end{equation}
This will  ensure that the sum of the GH charges of the solution
will remain  the same after the spectral flow.   After the gauge
transformation, the harmonic functions take the following form:
\begin{eqnarray}
V &=& \epsilon_0 ~+~ \ds\frac{1}{r}\,, \qquad K^1 ~=~ K^2 ~=~0\,,
\qquad K^{3} ~=~ k_0^3 ~+~ c\, \epsilon_0 ~+~ \ds\frac{c}{r} ~-~
\ds\sum_{i=1}^{N}\, \frac{q_3^i}{2 r_i}\,,
\nonumber\\
L_{1} &=& l_0^1 ~+~ \ds\sum_{i=1}^{N}\, \frac{Q_1^i}{4r_i}\,,
\qquad L_{2} ~=~ l_0^2 ~+~ \ds\sum_{i=1}^{N}\,
\frac{Q_2^i}{4r_i}\,,\qquad
L_{3} ~=~ l_0^3 \label{shiftharm}\\
M &=& m_0 ~-~\ds\frac{c\, l_0^3}{2} ~+~ \ds\sum_{i=1}^{N}\,
\frac{J_i}{16r_i}\,. \nonumber
\end{eqnarray}
To transform the solution corresponding to many supertubes to a
bubbling solution with an ambipolar Gibbons-Hawking base, we
perform a spectral flow transformation (\ref{GHspecflow1}) with
parameter $\gamma$ to obtain.
\begin{eqnarray}
\widetilde{V} &=& V ~+~ \gamma K_3\,, \qquad \widetilde{K}^{1} ~=~ K^{1} ~-~ \gamma L_2 \,, \qquad \widetilde{K}^{2} ~=~ K^{2} ~-~ \gamma L_1\,, \qquad \widetilde{K}^{3} ~=~ K^{3} \,,\nonumber  \\
\widetilde{L}_1 &=& L_1 \,, \qquad \widetilde{L}_2 ~=~ L_2 \,,
\qquad \widetilde{L}_3 ~=~ L_3 ~-~ 2\gamma M\,, \qquad
\widetilde{M} ~=~ M\,. \label{GHspecflow2}
\end{eqnarray}
The GH base space of the transformed solution has $N+1$ centers.
The new harmonic functions:
\begin{eqnarray}
\widetilde{V} &=& \tilde{\epsilon}_0 ~+~
\ds\sum_{j=1}^{N+1}\ds\frac{\tilde{q}_j}{r_j}~, \qquad
\widetilde{K}^{I} ~=~ \tilde{k}_0^I ~+~
\ds\sum_{j=1}^{N+1}\ds\frac{\tilde{k}_j^I}{r_j}~,
\\
\widetilde{L}^{I} &=& \tilde{l}_0^I ~+~
\ds\sum_{j=1}^{N+1}\ds\frac{\tilde{l}_j^I}{r_j}~,\qquad
\widetilde{M}~=~ \tilde{m}_0 ~+~
\ds\sum_{j=1}^{N+1}\ds\frac{\tilde{m}_j}{r_j}~,\nonumber
\end{eqnarray}
can be found straightforwardly from (\ref{shiftharm}) and
(\ref{GHspecflow2}). It is also straightforward to check that
(\ref{supermanyregular}), which insures the regularity of the
supertubes, implies that the constants in these harmonic functions
satisfy (\ref{chgchoice}) for any value of $\gamma$. Moreover, the
bubble equations (\ref{bubbleeqns}) are equivalent to the $N+1$
equations (\ref{radm0}) that give the radii of the $N$ supertubes
and the value of the $m_0$ parameter. This establishes explicitly
that for any even integer $\gamma $ the spectral flow
transformation (\ref{GHspecflow2}) maps smooth solutions
containing supertubes to smooth multi-center GH bubbling
solutions.

\subsection{Supertubes from Bubbling Geometries}

Having shown that a solution corresponding to many concentric
supertubes can be transformed into a GH bubbling solutions, it is
interesting to investigate the opposite transformation - that of a
bubbling solution into a solution containing supertubes.

It is not hard to see that given a generic smooth bubbling
solution, whose parameters respect (\ref{chgchoice}) and
(\ref{bubbleeqns}), one can perform a spectral flow
(\ref{GHspecflow1}) with parameter $\displaystyle \gamma = -{q_i
\over k^3_i }$ to obtain a solution in which there is no GH charge
at the $i^{\rm th}$ point. Equations (\ref{chgchoice}) then insure
that the functions $K^1,K^2$ and $L_3$ will also not have a pole
at the position of the $i^{\rm th}$ point. The poles of the other
harmonic functions are
\be 
K_3 \sim {k_i^3 \over r_i}~,~~~ L_1 \sim {- k_i^2 k_i^3 \over
r_i}~,~~~ L_2 \sim {-k_i^2 k_i^3 \over r_i}~,~~~ M  \sim {k_i^1
k_i^2 k_i^3\over 2 r_i}~. 
\ee
This solution corresponds to an object with two charges, one
dipole charge, and angular momentum, and it is simply a
circular\footnote{The circle is along the $U(1)$ fiber of the
ambi-polar Taub-NUT base.} two-charge supertube at position $\vec
r_i$.

It is clear that this solution will be smooth from a
six-dimensional perspective, simply because spectral flow takes
smooth solutions into smooth solutions. Moreover, the coefficients
of the singular parts of $L_1, L_2, K_3$ and $M$ satisfy the same
relation, (\ref{regularity}), as do the coefficients in the smooth two-charge
supertube solutions in $\IR^4$ or Taub-NUT. In
upcoming work \cite{inprogress} we will show that the smoothness
conditions coming from the supergravity analysis coincide with the
equations of motion for a two-charge supertube in a GH background
that one obtains using the Born-Infeld action of the supertube.
Hence, a spectral flow transformation with a well-chosen parameter
can transform \textit{any} multi-center Gibbons-Hawking solution
to a solution where one (or several) of the centers has been
replaced by a two-charge supertube.

\section{Generalized Spectral Flow}

It is also interesting to consider generalized spectral flow
transformations that can take a GH center into an even simpler
configuration. We begin by exploring the orbit of generalized
spectral flow.  We then use the physics of supertubes to argue
that many multi-center configurations that appear to be bound
states from a four-dimensional perspective do so only because of
the limited supergravity Ansatz used to study their stability.
When one explores them using a more complete supergravity Ansatz,
based on the underlying holographic dual, they are in fact
unbound.

\subsection{Multi-center D6-D4-D2-D0 Configurations}

In order to describe generalized spectral flow on multi-center
solutions it is convenient to work in the five-dimensional duality
frame in which the electric charges of the solution are those of
three sets of M2 branes (\ref{elevenmetric}). When the base space
of these solutions is ambi-polar, multi-center Taub-NUT, they can
be reduced to four-dimensional multi-center solutions. The M2
charges correspond to D2 charges, the M5 dipole charges correspond
to D4 charges,  the Kaluza-Klein momentum along the Taub-NUT fiber
becomes the D0 charge and the geometric GH charges correspond to
D6 branes. The sources that appear in the eight harmonic functions
that determine the solutions thus  correspond exactly to the
four-dimensional D6, D4, D2 and D0 charges.

A smooth multi-center, five-dimensional solution corresponds, in
four dimensions, to a multi-center solution where each center is
a ``primitive'' D6 brane, that is, a D6 brane that has non-trivial world-volume flux and
locally preserves sixteen supercharges\footnote{Only four of those supercharges are
common to all the D6 branes, and thus common to the complete multi-center
solution.}.  From the perspective of the D-brane world-volume, primitivity
places non-trivial constraints upon the fluxes. In the supergravity background
these constraints amount to imposing  smoothness, which fixes the flux parameters as in
(\ref{chgchoice}) \cite{Bena:2005va, Berglund:2005vb, Gimon:2007ps}.
In the same manner, a two-charge supertube, which is also smooth in the
D1-D5-P duality frame, has D4, D2 and D0 charges that satisfy
(\ref{regularity}). Thus it corresponds to a ``primitive'' D4
brane - a D4 brane with non-trivial world-volume flux that
locally preserves sixteen supercharges.

In section 3 we have established that spectral flow generically
takes multi-center, primitive D6 configurations into other such
configurations.  Moreover, in section 4 we have seen that for some
specially-chosen parameters it can transform a primitive D6 center
into a primitive D4 center. One can take this further, and
consider two- and three-parameter generalized spectral flow. A
two-parameter spectral flow, with
\be \gamma_1 = -{q \over k^1}~,~~~\gamma_2 = -{q \over k^2}
\label{twogamma} \ee
can take a GH center with GH charge $q$ into a center with only
two singular harmonic functions, $K_3$ and $M$. This corresponds
to a set of primitive D2 branes that have a non-trivial D0 brane
charge. Furthermore, one can perform another spectral flow to take
this center into a center that only has a non-zero $M$, and hence
corresponds to a collection of D0 branes. The parameters of this
flow are:
\be \gamma_1 = -{q \over k^1}~,~~~\gamma_2 = -{q \over
k^2}~,~~~\gamma_3 = -{q \over k^3}  \,. \label{threegamma}
\ee
Since this last configuration consists of only a single species of D-brane,
primitivity (the local preservation of sixteen supercharges) is now manifest.
One should note that each successive spectral flow decreases the number
of types of D-brane charge possessed by the brane and that this reduction critically depends
upon the selection of parameters, (\ref{chgchoice}), that made the original
fluxed D6-brane smooth.   By reversing these multiple spectral flows one
thus obtains another way to understand the primitivity of the original D6-brane configuration.

Unlike the primitive D6 branes, which give smooth five-dimensional
solutions in all duality frames, or the primitive D4 branes, which
are smooth in the D1-D5-P frame, the primitive D2 and D0 branes
are not smooth in supergravity in any duality frame.
This is not unexpected, because the U-duality group in four
dimensions can take smooth solutions into singular ones, and
generalized spectral flow is nothing but a three-parameter family
of this group. This is also not in conflict with the fact that
each spectral flow can be realized by a six-dimensional coordinate
change and therefore will preserve the regularity of
six-dimensional solutions. The point is that one cannot realize
all three independent spectral flows as coordinate changes of a
single regular metric and so concatenating spectral flows can
generate singular solutions.

One can also extend spectral flow to $U(1)^N$ five-dimensional
ungauged supergravities compactified on a GH space (or, after
dimensional reduction, $\mathcal{N}=2$ supergravities in four
dimensions), that come from M-theory compactified on a CY
manifold. The equation that gives generalized spectral flow
(\ref{GHspecflow}) is written in a way that trivially expands to
such supergravities\footnote{For a general CY compactification
$C_{IJK}$ are the triple intersection numbers of the CY
three-fold.}. For such solutions a six-dimensional lift of the
solution, and the smooth supertube interpretation of the primitive
D4 centers are not straightforward (unless the CY is $K^3 \times
T^2$). Nevertheless, for generic parameters the generalized
spectral flow still takes smooth solutions into smooth solutions,
while for special choices of parameters it can interpolate between
solutions with primitive D6, D4, D2 and D0 centers.

To recapitulate, from a five-dimensional perspective a spectral
flow with one parameter can take a smooth GH solution into a
smooth solution that contains a two-charge supertube in a GH
background. Furthermore, two-parameter and three-parameter
generalized spectral flows can transform a GH center into a
singular configuration, that from a four-dimensional perspective
has D2-D0 and pure D0 charges respectively.

\subsection{When is a multi-center solution a bound state?}

In studying the microstates of a black-hole one obviously wants to
ensure that one is studying a single black hole and not merely an
ensemble, or gas, of  unbound BPS black holes and black rings.
That is, one should only consider a system as being a single black
hole if there are no ``separation moduli'' that can be used to
physically deform the system into widely  separated components
without changing the energy or other asymptotic charges.

Establishing whether a solution is bound or unbound can be rather
subtle. Consider for example an asymptotically-flat
five-dimensional solution containing two two-charge supertubes.
These do not interact with each other, and one can move them
arbitrary far apart at no energy cost (without affecting the
asymptotic charges and angular momenta). Hence, this configuration
has flat directions, and is unbound. However, when considered as a
four-dimensional multi-center solution, this solution has three
centers that have a nontrivial four-dimensional $\vec E \times
\vec B$ interaction, and appears bound. Of course, the answer to
this puzzle is that the separation moduli of the five-dimensional
solution break its tri-holomorphic $U(1)$ isometry, and hence are
not visible in four dimensions.

If one's purpose is to describe microstates of five-dimensional
three-charge black holes or black rings, the ultimate arbiter of
whether a multi-center solution is bound is to dualize it to the
D1-D5-P duality frame, and take the limit in which it becomes
asymptotic to $AdS_3 \times S^3$ \cite{inprogress,microBR}. If the
six-dimensional solution has separation moduli, the solution is
not bound. As we will see below, these separation modes are often
not visible if one constructs and analyzes the solution using a
more limited four- or five-dimensional Ansatz\footnote{It would be
interesting to explore if the unbound nature of certain
multi-center solutions is also visible when they are embedded in
asymptotically $AdS_3 \times S^2$ solutions \cite{AdS3-S2}.}.

It is also possible that a certain multi-center solution, which is
unbound when embedded in an asymptotically $AdS_3 \times S^3$
spacetime, can become bound when embedded in an asymptotically
$\IR^{3,1} \times S^1 \times S^1 $ spacetime. The simplest example
is again that of two concentric two-charge supertubes in Taub-NUT.
These supertubes have no zero-modes \cite{srivastava} because of
the constraining nature of the Taub-NUT geometry. However, when
taking the limit in which their solution becomes asymptotically
$AdS_3 \times S^3$, the base of the solution becomes $\IR^4$, and
the two supertubes become indistinguishable from two supertubes in
an asymptotically-flat five-dimensional space, which are unbound.
The same analysis extends trivially to more concentric supertubes
in Taub-NUT.

Hence, two or more supertubes in Taub-NUT do not form a true bound
state. Rather they are {\it geometrically bound}: their lack of
separation moduli is a result of the compactification geometry
rather than of binding interactions.    Intuitively, one should
think of such geometrically-bound configurations as being the
analogue of an ideal gas in a box: there is no binding energy
between the atoms, but the system cannot be deformed into widely
separated components because of the walls of the box.

As we have seen in section 4, using spectral flow one can
transform a solution that contains concentric two-charge
supertubes to a bubbling multi-center solution. The analysis above
implies that such a bubbling multi-center solution  is not a bound state.
Indeed, upon spectral flow, a
solution where the supertubes are not concentric anymore becomes a
$v$-dependent six-dimensional solution (\ref{6Dmet}) of the type
described in section 2.2 \cite{Gutowski:2003rg,Cariglia:2004kk}.
Hence, as explained above, multi-center bubbling solutions that
can be obtained by spectral flow from a concentric-supertube
configuration only appear bound from a four- and five-dimensional
perspective because of the limited supergravity Ansatz used to
describe them. Upon embedding it in the correct holographic
background this configuration has at least one zero-mode and
this involves  making the six-dimensional lift of the solution
(\ref{6Dmet}) $v$-dependent.

It is also important to realize that starting from concentric
supertubes, a
 spectral flow transformation only
generates a very specific type of bubbling geometries. Indeed,
spectral flow leaves the bubble equations\footnote{Or the
``integrability conditions" in the case of more general centers
\cite{denef1}.} invariant, and hence the bubble equations
governing the unbound bubbling solutions do not contain any terms
that depend on the distance between any two of the  GH centers
that come from the supertubes. Hence, from a four-dimensional
perspective these GH points are free to move on two-spheres of
radius, $R_i$ (given by (\ref{radm0})) around a central GH point.
In the quiver language, used to describe multi-center
four-dimensional solutions \cite{denef1}, these bubbling solutions
can be depicted as ``hedgehog'' quivers, with all the arrows
originating from one of the nodes, and joining it to all the other
nodes.

\begin{figure}
\centering
\includegraphics[height=6cm]{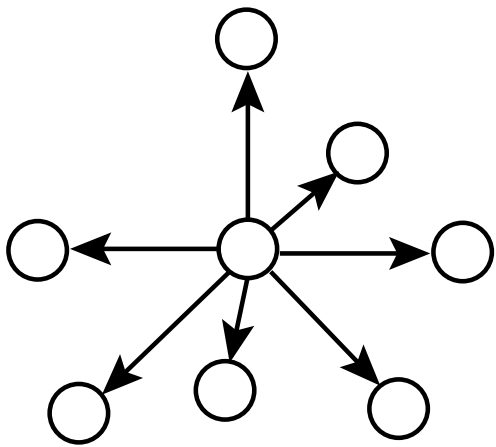}
\caption{The ``hedgehog'' quiver corresponding to the unbound
multi-center
  solution generated from the spectral flow of many concentric
  supertubes. }
\label{GHquiver}
\end{figure}

It is also possible to argue that, at least for a large enough
number of centers, spectral flow can be used to generate {\it all}
the hedgehog multi-center GH solutions in Figure (\ref{GHquiver})
from two-charge round supertubes in flat space. Indeed, by simple
parameter counting, a solution with $N+1$ GH centers has $4N+1$
parameters (three $k^I_i$'s  and one GH charge $q_i$ for each
point minus three gauge transformations). Requiring the vanishing
of $\Pi^{(1)}_{ij} \Pi^{(2)}_{ij} \Pi^{(3)}_{ij}$ between any two
of $N$ centers naively imposes $N(N-1)$ constraints, which, in
general, cannot be satisfied. Nevertheless, since $
\Pi^{(I)}_{ij}$ are given by (\ref{chgredefs}), it is not hard to
see that one can also have all of them zero if for one of the $I$,
the value of $k^I_i/q_i$ is the same for all the N points.
Choosing, for example, $I=2$, this implies
\be k_i = (k^1_i, q_i \kappa^2 , k_i^3)~, \label{equalflux} \ee
and imposes  $N-1$ conditions, leaving $3N+2$ independent
parameters. This is exactly the number of parameters that describe
all possible spectral flows of $N$ round supertubes of arbitrary
charge in a Taub-NUT space:  three independent parameters
$(Q_1,Q_2,d_3)$ for each supertube, one for the Taub-NUT center
and one spectral flow parameter, $\gamma$.   It is not hard to see
that a spectral flow with parameter $\gamma_2 = - 1/\kappa^2 $
transforms the foregoing set of $N$ GH centers into concentric
two-charge supertubes.

There exists another way to make all the fluxes between the $N$ GH
points vanish: one can divide them in three sets, $A,B,C$, that
have fluxes
\bea
k_i &=& (k^1_i,q_i \kappa^2 , q_i \kappa^3)~,~~~ {\rm for}~ i \in A  \nonumber \\
k_i &=& (q_i \kappa^1,k^2_i,q_i \kappa^3)~,~~~ {\rm for}~ i \in B   \label{funnyflux} \\
k_i &=& (q_i \kappa^1, q_i \kappa^2, k^3_i)~,~~~ {\rm for}~ i \in
C  \nonumber \eea
where $\kappa^1, \kappa^2, \kappa^3$ are constants, and
$k^1_i,k^2_i$ and $k^3_i$ can be arbitrary for the GH centers in
the $A,B$ and $C$ set respectively. A two-parameter spectral flow
with parameters $\gamma_1 = -1/\kappa^1$ and $\gamma_2 =
-1/\kappa^2$ transforms the GH centers in the $A$ and $B$ sets
into two-charge supertubes of different type, and transforms the
GH centers in the $C$ set into singular D2-D0 centers. Normally,
different kinds of two-charge supertubes have an $\vec E \times
\vec B$-type electric-magnetic interaction, and cannot go
arbitrarily far away from each other without changing the
asymptotic charges of the solution. Therefore such a solution is
generically a bound state. However, for the particular supertubes
that are created from the spectral flow of (\ref{funnyflux}) the
$\vec E \times \vec B$-type interaction vanishes, and hence they
can move freely away from each other. Hence this type of
configuration also corresponds to an unbound state. It is quite
clear that for  $N$ sufficiently large all hedgehog quivers can be
only of the type (\ref{equalflux}) or (\ref{funnyflux}), and hence
they are all unbound from the point of view of six-dimensional
supergravity.

Another intuitive way to think of our formulation of bound state
classification is as follows.  Bound states generically emerge
through $\vec E \times \vec B$ interactions.  In four dimensions
there are four independent $U(1)$ Maxwell fields and thus many
ways in which to generate the interaction.   In five dimensions
there are only three $U(1)$ Maxwell fields and thus some of the
four-dimensional $\vec E \times \vec B$ interactions become
trivial upon oxidation to five dimensions.  Indeed, this is
precisely what happens  with the hedgehog quiver:  One can map the
sources of the four-dimensional $\vec E \times \vec B$  interaction to a single
D6 at the center of the quiver and D0 charges on the nodes.  Upon lifting to five
dimensions, the D6 brane disappears (becoming the  ``center of space'') and all the
nodes become free.

One should note that our analysis here indicates that the hedgehog quiver
describes unbound states only when the center of the quiver is primitive. Indeed, one can
consider quivers in which the exterior nodes have charges
corresponding to  black rings, and the center is a primitive (fluxed) D6
brane\footnote{One could also consider spectrally-flowed version
of this configuration, in which the D6 brane is transformed into
another primitive brane.}. This solution can be lifted to one or
many concentric black rings on an $\IR^4$ base. As with
supertubes, the absence of arrows between the exterior nodes of
the quiver is equivalent to the absence of $\vec E \times \vec
B$-like interactions between the black rings. Hence, in the
asymptotically five-dimensional solution, these rings can slide
away from each other, and the configuration has zero modes and is
unbound.  However, if the center node is not a fluxed D6 brane,
but a BMPV black hole, the sliding away of the rings becomes
impossible. Indeed, as shown in \cite{offcenter}, one cannot take
a BMPV black hole away from the center of a black ring without
modifying the asymptotic angular momenta of the solution. In that
case, the black ring and the black hole interact via  $\vec  E
\times \vec B$-type interactions, that render the sliding-away
mode massive.

To summarize, our arguments indicate that all asymptotically
five-dimensional solutions given by hedgehog quivers  are unbound when
their centers are primitive branes. Although a more detailed
analysis is needed, this also seems to be true for hedgehog
quivers whose central node is primitive and the outside nodes
are not. However, the system may well be a bound
state when the central node is not primitive.

A few examples of quivers describing unbound states include, for
example the ``Hall halo'' configurations with a primitive center
discussed in \cite{denef1}, the three-center configuration
discussed in Section 6 of \cite{Berglund:2005vb}, and possibly
also the ``foaming quiver'' (with charges equal to those of a
maximally-spinning BMPV black hole) considered in
\cite{Bena:2006is}.  As we stressed earlier, the unbound status of
such geometrically bound systems can only be seen when considering
asymptotically-flat five-dimensional solutions, or asymptotically
$AdS_3 \times S^3$ solutions in six dimensions.

This analysis of bound and unbound systems is also in agreement
with the recent findings that only quivers with closed loops can
give solutions that have the charges of black holes and black
rings with classically large horizon radius
\cite{Bena:2006kb,abyss}, and that at weak coupling only these
quivers give a macroscopic (black-hole-like) entropy
\cite{denefmoore}.  Based on this, one expects that the closed,
deep or scaling quivers describe bound states, which they indeed
do \cite{Bena:2006kb}.  Intuitively, one can think about the microstates that come
from hedgehog quivers (and are necessarily ``shallow"), and
possibly about other ``shallow'' microstates as unbound or
very weakly bound; conversely, the  deep  $AdS$ throat, which is
the hallmark of the scaling solution, is a direct manifestation of
the binding of the geometric components of the microstate
geometry.

\section{Conclusions}

We have investigated spectral flow - a transformation that takes
multi-center solutions into other multi-center solutions, by
shifting the underlying harmonic functions. For generic
parameters, this transformation takes bubbling solutions that have
multiple GH centers into other bubbling solutions with GH centers.
However, for specially-chosen parameters, a spectral flow
transformation can take a bubbling solution into a smooth solution
that contains one or several supertubes in a GH multi-center
background.

The fact that spectral flow can be used to interchange solutions
containing two-charge supertubes and bubbling solutions is a very
powerful fact, which we have exploited throughout this paper, and
will continue to use in upcoming work.

The first problem we have addressed using this tool is to
understand which of the three-charge bubbling solutions
constructed in the literature are bound states, and which are not.
We have found that the solutions that correspond to quivers
without loops or bifurcations  can be transformed into solutions
that in the five-dimensional lift can be taken apart. Therefore
they should not correspond to bound states in the dual CFT.

The second use of spectral flow has been to prove the existence of
three-charge smooth BPS solutions that depend on arbitrary
functions in the vicinity of {\it any} multi-center GH solution.
Indeed, any GH center can be related to a round supertube via
spectral flow. Furthermore, supertubes can have arbitrary shape
while still remaining regular and supersymmetric. Hence the
inverse spectral flow of a wiggly supertube gives a new smooth
black-hole microstate solution, that does not have a GH base, and
that can depend on {\it arbitrary functions}.

Even without knowing the explicit form of these BPS solutions, it
is still possible to investigate their physics (at least in the
vicinity of GH solutions), analyze their moduli space, and count
their entropy by using supertube counting techniques
\cite{Palmer:2004gu,Bak,Slava}. The fact that GH solutions can be
deformed to BPS solutions that depend on arbitrary functions
establishes the existence of families of black hole microstates
that depend on an infinite number of continuous parameters. In
upcoming work \cite{entropycount} we will explore these
microstates, and argue that they can have a macroscopically large
(black-hole-like) entropy.

We have also explored a larger class of spectral flow
transformations (called generalized spectral flow) that for
generic parameters transform multi-center bubbling solutions into
other multi-center bubbling solutions, but for special parameters
can transform one or several of the centers of a bubbling solution
into a two-charge supertube, D2-D0 or D0 center.

By taking the limit of parameters in which the D2-D0, or the D0
branes do not back-react on the geometry, one can study them using
their (non-abelian) Born-Infeld action, and count their entropy. In fact,
such a counting has been performed in
several circumstances. For example, in
\cite{deconstruction,deconstruction-eric} it was found that the
entropy coming from D0 branes in a D6-$\rm \overline{D6}$
background (which lifts to a multi-center GH solution in five
dimensions) is of the same order as the would-be black hole
entropy. One could then use generalized spectral flow to transform
the D0-D6-$\rm \overline{D6}$ system considered there into a
multi-center D6-$\rm \overline{D6}$ configuration, which (unlike
the system with D0 branes) is well-described by supergravity in
the regime of parameters where the classical black hole exists. It
would be very interesting to follow the spectral flow, and find
the description of the D0 configurations that give the
black-hole-like entropy \cite{deconstruction,deconstruction-eric}
in the multi-center D6 frame, and to find whether these
configurations are still smooth in supergravity.

Moreover, in \cite{denefmoore} the entropy of a three-center
D6-D6-$\rm \overline{D6}$ configuration was computed at
intermediate coupling (using the quiver quantum mechanics
describing these branes) and found to give a macroscopic,
black-hole-like answer when the branes form a scaling solution. It
would be interesting to spectrally flow one of the D6 branes into
a two-charge supertube, and to see if the modes that give the
macroscopic entropy of the D6-D6-$\rm \overline{D6}$ are the same
as the fluctuation modes of the supertube. This would also help
find a realization of smooth BPS black hole microstates in the
regime of parameters where the black hole exists.

\bigskip
\leftline{\bf Acknowledgements}
\smallskip
We would like to thank Stefano Giusto, Samir Mathur and Ashish
Saxena for interesting discussions. NB and NPW are grateful to the
SPhT, CEA-Saclay for kind hospitality while this work was
completed. The work of NB and NPW was supported in part by funds
provided by the DOE under grant DE-FG03-84ER-40168. The work of IB
was supported in part by the \textit{Dir\'ection des Sciences de
la Mati\`ere} of the \textit{Commissariat \`a L'En\'ergie
Atomique} of France, by the French \textit{Agence Nationale de la
Recherche}, and by a Marie Curie International Reintegration
Grant. The work of NB was also supported by the John Stauffer
Fellowship from USC and the Dean Joan M. Schaefer Research
Scholarship.

\section*{Appendix A. $SL(2,\ZZ)^3$ transformations}
\appendix
\renewcommand{\theequation}{A.\arabic{equation}}
\setcounter{equation}{0}  

If one applies three $SL(2,\ZZ)$ transformations (\ref{sl2zxy}) on
the three pairs formed by $x_{78}$ and $ y_{12}$, $ y_{34}$ and $
y_{56}$ respectively (and permutations thereof), one obtains
\begin{equation}
\widetilde{X} ~=~ (B_3^{-1} A_3 B_3 B_2^{-1} A_2 B_2 B_1^{-1} A_1
B_1) \cdot X
\end{equation}
where
\begin{equation}
\widetilde{X} \equiv \{ \tilde{x}_{12}, \tilde{x}_{34},
\tilde{x}_{56}, \tilde{x}_{78}, \tilde{y}_{12}, \tilde{y}_{34},
\tilde{y}_{56}, \tilde{y}_{78} \}, \qquad X \equiv \{ x_{12},
x_{34}, x_{56}, x_{78}, y_{12}, y_{34}, y_{56}, y_{78} \}
\end{equation}
and
\begin{equation}
B_1 = \text{diag}(1,2,2,1,2,1,1,2)\, \qquad B_2 =
\text{diag}(2,1,2,1,1,2,1,2)\, \qquad B_3 =
\text{diag}(2,2,1,1,1,1,2,2).
\end{equation}
The diagonal matrices $B_I$ fix the correct periodicity for the
$\tau$ and $v$ coordinates, namely $0\leq\tau< 4\pi$ and $0\leq v<
2\pi$. The matrices $A_I$ are three commuting, rank 8 matrices
composed of the matrix elements of the three  $SL(2,\ZZ)$ matrices
$\mathcal{M}_I$. They have the explicit form
\begin{equation}
A_1 = \left(\begin{array}{cccccccc}
m_1 & 0 & 0 & 0 & 0 & 0 & 0 & n_1\\
0 & q_1 & 0 & 0 & 0 & 0 & p_1 & 0\\
0 & 0 & q_1 & 0 & 0 & p_1 & 0 & 0\\
0 & 0 & 0 & m_1 & n_1 & 0 & 0 & 0\\
0 & 0 & 0 & p_1 & q_1 & 0 & 0 & 0\\
0 & 0 & n_1 & 0 & 0 & m_1 & 0 & 0\\
0 & n_1 & 0 & 0 & 0 & 0 & m_1 & 0\\
p_1 & 0 & 0 & 0 & 0 & 0 & 0 & q_1
\end{array}\right)
\end{equation}
\begin{equation}
A_2 = \left(\begin{array}{cccccccc}
q_2 & 0 & 0 & 0 & 0 & 0 & p_2 & 0\\
0 & m_2 & 0 & 0 & 0 & 0 & 0 & n_2\\
0 & 0 & q_2 & 0 & p_2 & 0 & 0 & 0\\
0 & 0 & 0 & m_2 & 0 & n_2 & 0 & 0\\
0 & 0 & n_2 & 0 & m_2 & 0 & 0 & 0\\
0 & 0 & 0 & p_2 & 0 & q_2 & 0 & 0\\
n_2 & 0 & 0 & 0 & 0 & 0 & m_2 & 0\\
0 & p_2 & 0 & 0 & 0 & 0 & 0 & q_2
\end{array}\right)
\end{equation}
\begin{equation}
A_3 = \left(\begin{array}{cccccccc}
q_3 & 0 & 0 & 0 & 0 & p_3 & 0 & 0\\
0 & q_3 & 0 & 0 & p_3 & 0 & 0 & 0\\
0 & 0 & m_3 & 0 & 0 & 0 & 0 & n_3\\
0 & 0 & 0 & m_3 & 0 & 0 & n_3 & 0\\
0 & n_3 & 0 & 0 & m_3 & 0 & 0 & 0\\
n_3 & 0 & 0 & 0 & 0 & m_3 & 0 & 0\\
0 & 0 & 0 & p_3 & 0 & 0 & q_3 & 0\\
0 & 0 & p_3 & 0 & 0 & 0 & 0 & q_3
\end{array}\right)
\end{equation}
Note that
\begin{equation}
\text{Det}A_{I} = (m_Iq_I - n_Ip_I)^4 = 1, \qquad \text{and}
\qquad \text{Tr}A_I = 4 (m_I+q_I) = 4 \text{Tr}\mathcal{M}_I
\end{equation}
and also
\begin{equation}
\text{Tr}(A_1A_2A_3) = (m_1 + q_1)(m_2+ q_2)(m_3 + q_3) =
\ds\frac{1}{64} (\text{Tr}A_1)(\text{Tr}A_2)(\text{Tr}A_3) \,.
\end{equation}
Thus, in general, the new solution obtained after spectral flow is
determined by the eight harmonic functions $\{ \widetilde{V},
\widetilde{K}_{I}, \widetilde{L}_{I}, \widetilde{M} \}$ each of
which is a  linear combination of the eight
harmonic functions that determine the original solution $\{ V,
K_{I}, L_{I}, M \}$.

For non-zero $m_I$ and $q_I$ one can  use the gauge
transformations (\ref{gaugetransf}), which leave the physical
solution invariant, to set $p_1=p_2=p_3=0$. Then the most general
transformation on the eight harmonic functions is:
\begin{eqnarray}
\widetilde{L}_{I} &=&
\ds\frac{1}{2}\delta_{IJ}C^{JKL}m_{J}q_{K}q_{L}L_{J} +
2\delta_{IJ} C^{JKL}n_Jq_Kq_L M,\qquad \widetilde{M} =
\ds\frac{1}{6} C^{JKL}q_Jq_Kq_L M, \qquad
\vec{\tilde{\omega}} = \vec{\omega} \nonumber\\ \nonumber\\
\widetilde{K}^{I} &=&
\ds\frac{1}{2}\delta^{I}_{J}C^{JKL}q_Jm_Km_LK^J +
2\delta^{I}_{J}C^{JKL}q_Jm_Kn_LL_K +
4\delta^{I}_{J}C^{JKL}q_Jn_Kn_LM \label{sl2zgen}
\\ \nonumber\\
\widetilde{V} &=& \ds\frac{1}{6}C^{JKL}m_Jm_Km_L V -
C^{JKL}n_Jm_Km_LK^J - 2C^{JKL}m_Jn_Kn_LL^J - \ds\frac{8}{3}
C^{JKL}n_Jn_Kn_LM\nonumber
\end{eqnarray}
For $m_I = q_I = 1$ one obtains the generalized spectral flow
transformation (\ref{GHspecflow}) (using $\gamma_I = - 2n_I$).

For non-trivial transformations with $m_I=0$ or $q_I=0$, one
has $n_I=-p_I=\pm1$, and one cannot use the gauge transformation
(\ref{gaugetransf}) to set $p_1=p_2=p_3=0$.  Hence the $SL(2,\ZZ)^3$
transformation does {\it not} reduce to a generalized spectral
flow. For example, when both $m_I=0$ and $q_I=0$, the new harmonic
functions are:
\begin{eqnarray}
\widetilde{V} &=& - \ds\frac{8}{3} C^{IJK}n_In_Jn_K M = \pm 16 M,
\qquad\qquad \widetilde{M} = - \ds\frac{1}{12}
C^{IJK}\ds\frac{p_Ip_Jp_K}{8}
V = \pm \ds\frac{1}{16} V\,, \nonumber\\\\
\widetilde{L}_I &=& - \ds\frac{1}{2}\delta_{IJ}
C^{JKL}\ds\frac{n_Jp_Kp_L}{2} K^J = \pm \ds\frac{1}{2} K^I,
\qquad\qquad \widetilde{K}^I = - \ds\frac{1}{2} \delta_{IJ}
C^{JKL} 2p_Jn_Kn_L L_J = \pm 2 L_I \,,\nonumber
\end{eqnarray}
which is  an interchange of the $x$ and $y$ parameters of
$E_{7(7)}$, and in four dimensions corresponds to an
electric-magnetic duality.

It is also interesting to note that although the transformations
(\ref{sl2zgen}) are obtained for $U(1)^3$ ungauged supergravity in
five dimensions, they are also valid for a more general $U(1)^N$
supergravity obtained from M-theory compactified on a Calabi-Yau
3-fold.

One may consider also other $SL(2,\mathbb{Z})$ transformations on
the harmonic functions $x_{a,a+1}$, $y_{a,a+1}$. However these
will not preserve the regularity of the initial bubbling solution.
Consider for example the transformation
\begin{equation}
\begin{array}{l}
\left( \begin{array}{c}
\tilde{x}_{12}\\
2\tilde{y}_{12}
\end{array} \right) = \mathcal{M}_4 \left( \begin{array}{c}
x_{12}\\
2y_{12}
\end{array} \right), \qquad\qquad \left( \begin{array}{c}
\tilde{x}_{34}\\
2\tilde{y}_{34}
\end{array} \right) = \mathcal{M}_4 \left( \begin{array}{c}
x_{34}\\
2y_{34}
\end{array} \right) \,, \\   \\
\left( \begin{array}{c}
\tilde{x}_{56}\\
2\tilde{y}_{56}
\end{array} \right) = \mathcal{M}_4 \left( \begin{array}{c}
x_{56}\\
2y_{56}
\end{array} \right), \qquad\qquad \left( \begin{array}{c}
\tilde{x}_{78}\\
2\tilde{y}_{78}
\end{array} \right) = \mathcal{M}_4 \left( \begin{array}{c}
x_{78}\\
2y_{78}
\end{array} \right) \,,
\end{array}
\label{sl2zxy2}
\end{equation}
which could be expressed as
\begin{equation}
\widetilde{X} ~=~ (B_4^{-1}A_4B_4) \cdot X  \,,
\end{equation}
where
\begin{equation}
B_4 = \text{diag} (1,1,1,1,2,2,2,2)
\end{equation}
and
\begin{equation}
A_4 = \left(\begin{array}{cccccccc}
m_4& 0 & 0 & 0 & n_4 & 0 & 0 & 0\\
0 & m_4 & 0 & 0 & 0 & n_4 & 0 & 0\\
0 & 0 & m_4 & 0 & 0 & 0 & n_4 & 0\\
0 & 0 & 0 & m_4 & 0 & 0 & 0 & n_4\\
p_4 & 0 & 0 & 0 & q_4 & 0 & 0 & 0\\
0 & p_4 & 0 & 0 & 0 & q_4 & 0 & 0\\
0 & 0 & p_4 & 0 & 0 & 0 & q_4 & 0\\
0 & 0 & 0 & p_4 & 0 & 0 & 0 & q_4
\end{array}\right) \,.
\end{equation}
As before we have the relations
\begin{equation}
\text{Det}A_{4} = (m_4q_4 - n_4p_4)^4 = 1, \qquad \text{and}
\qquad \text{Tr}A_4 = 4 (m_4+q_4) = 4 \text{Tr}\mathcal{M}_4~.
\end{equation}
Note that the matrix $A_4$ does {\it not} commute with the other
three matrices $A_I$. More importantly the new solution specified
by the eight harmonic functions $\tilde{x}_{a,a+1}$,
$\tilde{y}_{a,a+1}$ is not regular because it fails to satisfy the
regularity conditions:
\begin{equation}
\tilde{l}_i^{I} = - \ds\frac{1}{2} C_{IJK}
\ds\frac{\tilde{k}_i^J\tilde{k}_i^K}{\tilde{q}_i} \qquad
\text{and} \qquad \tilde{m}_i = \ds\frac{1}{2}
\ds\frac{\tilde{k}_i^1\tilde{k}_i^2\tilde{k}_i^3}{(\tilde{q}_i)^2} \,.
\end{equation}
One can, of course, apply other $SL(2,\mathbb{Z})$ transformations
that are subgroups of the $E_{7(7)}$ U-duality group of the
supergravity theory. However such transformations will generically
convert a regular bubbling solution to an singular one. It is
interesting to observe that the $SL(2,\mathbb{Z})$ transformation
which produced a singular solution ($\mathcal{M}_4$ above) leads
to an $8\times8$ matrix that does not commute with the three
matrices $A_I$ that correspond to the $SL(2,\mathbb{Z})^3$ that
generically takes smooth solutions into smooth solutions. One
could speculate that there might exist a relation between the fact
that the matrices $A_1, A_2$ and $A_3$ commute and the fact that
they generate the largest subgroup of $E_{7(7)}$ that preserves
smoothness, but we will leave the investigation of this to future
work.



\end{document}